# Enhanced Catalytic Activity in Strained Chemically Exfoliated WS$_2$ Nanosheets for Hydrogen Evolution

Damien Voiry*, Hisato Yamaguchi*, Junwen Li⌘, Rafael Silva+, Diego C B Alves*, Takeshi Fujita#§, Mingwei Chen#, Tewodros Asefa+||, Vivek Shenoy⌘, Goki Eda$,$$, and Manish Chhowalla*

*Rutgers University, Materials Science and Engineering, 607 Taylor Road, Piscataway, New Jersey 08854, USA.
⌘Brown University, School of Engineering, Providence, Rhode Island 02912, USA.
+Rutgers University, Department of Chemistry and Chemical Biology, 610 Taylor Road, Piscataway, New Jersey 08854, USA
#WPI Advanced Institute for Materials Research, Tohoku University, Sendai 980-8577, Japan.
§JST, PRESTO, 4-1-8 Honcho Kawaguchi, Saitama 332-0012, Japan
|| Rutgers University, Department of Chemical and Biochemical Engineering, 98 Brett Road, Piscataway, New Jersey 08854, USA.
$$National University of Singapore, Physics Department and Graphene Research Centre, 2 Science Drive 3, Singapore 117542.
$$National University of Singapore, Chemistry Department, 3 Science Drive 3, Singapore 117543.

**The ability to efficiently evolve hydrogen via electrocatalysis at low overpotentials holds tremendous promise for clean energy [1-10]. Hydrogen evolution reaction (HER, 2H$^+$ + 2e$^-$ → H$_2$) can be easily achieved from water if a voltage above the thermodynamic potential of the HER is applied. Large overpotentials are energetically inefficient but can be lowered with expensive platinum based catalysts. Replacement of Pt with inexpensive, earth abundant electrocatalysts would be significantly beneficial for clean and efficient hydrogen evolution. Towards this end, promising HER characteristics have been reported using 2H (trigonal prismatic) XS$_2$ (where X = Mo or W) nanoparticles with a high concentration of metallic edges [11] as electrocatalysts [3,4,6]. The key challenges for HER with XS$_2$ are increasing the number and catalytic activity of active sites [2]. Here we report atomically thin**

**nanosheets of chemically exfoliated WS$_2$ as efficient catalysts for hydrogen evolution with very low overpotentials. Atomic-resolution transmission electron microscopy and spectroscopy analyses indicate that enhanced electrocatalytic activity of WS$_2$ is associated with high concentration of strained metallic 1T (octahedral) phase in the as-exfoliated nanosheets. Density functional theory calculations reveal that the presence of strain in the 1T phase leads to an enhancement of the density of states at the Fermi level and increases the catalytic activity of the WS$_2$ nanosheet. Our results suggest that chemically exfoliated WS$_2$ nanosheets could be interesting catalysts for hydrogen evolution.**

Realization of the "hydrogen economy" will require efficient and sustainable production of hydrogen. The scientific community has been actively searching for new electrocatalysts that decrease the overpotential of the HER as well as being earth abundant and inexpensive [1,7,12-16]. Platinum is the most electroactive and electrochemically stable catalyst, making it challenging to find a suitable replacement. The identification of XS$_2$ as potential efficient catalysts for HER has opened up an exciting new path for the field [3-6,16,17]. Much of the experimental work has stemmed from theory that suggested that metallic edges [11] of 2H XS$_2$ crystals are electrocatalytically active [3]. This has led to investigations of XS$_2$ nanoparticles or complexes with high concentration of exposed edges as electrocatalysts for hydrogen evolution [5,15,16-19]. Further improvements in the HER kinetics by placing XS$_2$ nanoparticles on a variety of substrates including reduced graphene oxide have also been observed [20].

$XS_2$ belong to the layered transition metal dichalcogenides (LTMDs) family of compounds that have lamellar structure comparable to graphite in which the individual S-Metal-S layers weakly interact with each other. Similar to graphite, LTMDs can be exfoliated into their individual atomically thin sheets by mechanical, chemical, or electrochemical means [21-24]. Liquid based exfoliation of layered compounds has been known and studied over several decades but recently received renewed interest with the work of Coleman et al. [24]. In addition to chemical exfoliation, chemical vapor deposition of ultra-thin $XS_2$ layers has also been achieved [25-27]. Recently such CVD layers on nickel foams have been investigated for hydrogen evolution [28]. The ease of fabrication along with high carrier mobility [29] and interesting optical properties [23,30,31] make single layered LTMDs attractive for a wide variety of opto-electronic applications but their catalytic properties remain largely unexplored.

We have recently utilized a well-established chemical method [21] to efficiently exfoliate $WS_2$ into atomically thin platelets [32] and deposit uniform thin films from them. The as-exfoliated nanosheets predominantly consist of 1T structure that is metallic [33], in contrast to the semiconducting 2H phase (Supplementary Information). The 1T phase is metastable so that annealing of as-exfoliated sheets results in restoration of the 2H phase, which is accompanied by metallic to semiconducting electronic transition [34]. The band gap of the annealed 2H phase is ~ 2.0 eV, ideal for photocatalysis [35]. In this study, we have explored the electrocatalytic properties for HER of atomically thin $WS_2$ nanosheets.

To prepare the nanosheets, commercially available $WS_2$ powder was lithium intercalated to form $Li_xWS_2$, which can be readily exfoliated via forced hydration to form a stable suspension (Methods) [36]. The electrodes for HER were prepared by drop casting a fixed volume and concentration of $WS_2$ sheets from an aqueous suspension onto glassy carbon. A typical atomic force microscopy (AFM) image of exfoliated $WS_2$ nanosheets deposited by drop casting is shown in Figure 1a. Analyses of AFM images revealed that most flakes have lateral dimensions ranging from 100 – 800nm and thickness of ~ 1nm.

The structure of chemically as-exfoliated $XS_2$ has been studied in detail and it is agreed that it departs from an ideal 1T structure of $TiS_2$ [34]. Through a series of studies, Kanatzidis et al. [34,37] have concluded that the structure can be described as a highly distorted 1T structure with $2a_0 \times a_0$ superlattice. We have utilized high angle annular dark field (HAADF) imaging in an aberration corrected scanning transmission electron microscope (STEM) to demonstrate that the structure of as-exfoliated $WS_2$ consists primarily of disordered superlattice regions (Figure 1b). Upon annealing, the 1T phase relaxes to the stable 2H structure as shown in Figure 1c. The superlattice structure is the distorted and strained 1T phase that is well known to develop during Li intercalation [34]. The corresponding schematics of the distorted 1T and 2H phases are shown in the Supplementary Information (Figure S1) to help visualize the atomic structure. The STEM-HAADF image in Figure 1b is the first direct observation of the zigzag pattern that was previously inferred via spectroscopic and diffraction techniques for chemically exfoliated $MoS_2$ and $WS_2$ [37]. For the zigzag chain superlattice regions, we observe two distinct W-W distances, 2.7 Å and 3.3 Å that are substantially different from 3.15 Å for pristine 2H-WS$_2$.

The presence of the zig-zag structure in Figure 1b indicates a large concentration of locally strained bonds. We analyzed the images utilizing peak pairs analysis (PPA) [38] to obtain a strain tensor map of the 1T structure (Methods). The inset of Figure 1b shows the corresponding strain tensor map of mean dilatation [$\Delta_{ij} = ½ (\partial u_i/\partial x_i + \partial u_i/\partial x_j)$]. A complete stress tensor map is provided in the Supplementary Information (Figure S2). It can be seen that frequent occurrence of kinks in the zigzag chains leads to strong deformation and regions of high local strain which gives rise to an overall isotropic strain. The net change in the lattice parameter has been suggested to be ~ 3 % by X-ray diffraction analysis [39], which is comparable to what we have observed by STEM-HAADF. We have also correlated the STEM-HAADF observations with Raman and X-ray photoelectron spectroscopic (XPS) analyses, revealing the presence of the metallic 1T phase in the as-exfoliated nanosheets (Figure S3).

The HER with a monolayer of WS$_2$ nanosheet thin film as catalyst on glassy carbon was measured using the standard three-electrode electrochemical configuration in 0.5 M H$_2$SO$_4$ electrolyte de-aerated with Ar, as described in the Methods section. The polarization curves showing the normalized current density versus voltage (*j* vs *V*) for the 1T and 2H films along with Pt and bulk WS$_2$ powder samples, for comparison, are shown in Figure 2a. In addition, a sub-monolayered film of as-exfoliated WS$_2$ nanosheets is also shown in Figure 2a. It can be seen from the inset in Figure 2a that for the as-exfoliated $WS_2$ thin film electrodes the reaction begins at exceptionally low cathodic voltages ranging from 30 - 60 mV, above which the current increases rapidly. The overpotential and the Tafel slopes are slightly higher in the case of sub-monolayered $WS_2$ thin films, which is attributed to poor charge transfer from the film to the electrode [20]. That is below the percolation threshold, transport through the sheets is poor and thus the overall conductance is low which limits charge transfer and the catalytic activity. The 2H $WS_2$ electrodes show HER characteristics that are consistent with $MoS_2$ edges [4,6] with typical overpotentials of 200 – 300mV while the bulk powder exhibits poor catalytic activity. Additional insight into the catalytic activity of the $WS_2$ electrodes was obtained by extracting the slopes from Tafel plots shown in Figure 2b. The lowest Tafel slopes of ~ 60 mV/decade were obtained for the 1T $WS_2$ phase. These values are comparable to those measured for $MoS_2$ nanoparticles [4,18], suggesting that the surface chemistry mechanisms responsible for HER may be similar even though the geometry and composition are different. To quantify the catalytic activity, we measured the actual number of active sites using the under potential deposition method [40] (Figure S9, S10). Based on this method, we have determined the number of active sites to be ~ $4.5 \times 10^{14}$ sites/$cm^2$ or higher for the highly strained 1T phase. The turnover frequency (TOF) of $H_2$ molecules evolved per second (represented as $s^{-1}$) was calculated to be $175 s^{-1}$ at −288mV. The TOF as a function of the voltage for the different specimens tested in this study are plotted in Figure 2c. For comparison, the best TOF reported for molecular catalyst reach about $1000\ s^{-1}$ at overpotential of 290 mV and $90\ s^{-1}$ at 400 mV for Ni-based systems [41] and $MoS_2$-based systems respectively [15]. However such complexes are currently limited by their stability at low pHs and their assembly onto an electrode in order to operate in fully aqueous conditions. To this aim, Andreiadis et al. have recently reported overpotential around 350 mV at Tafel slope about 160 mV/dec when grafting Co complexes onto multiwalled carbon nanotubes as an electrode running in aqueous electrolyte [42]. Pt (111) has an exchange current density (by definition at 0V) of $4.5 \times 10^{-4}$ A/cm$^2$ giving a TOF of ~ $0.9s^{-1}$ while MoS$_2$ edges have exchange current densities of ~ $8 \times 10^{-6}$ A/cm$^2$ and TOF of around $0.02s^{-1}$ at pH of 0.24 [4]. In comparison, at 0.5 M of H$_2$SO$_4$, the exchange current density of exfoliated 1T WS$_2$ is ~ $2 \times 10^{-5}$ A/cm$^2$ and a TOF of ~ $0.043s^{-1}$.

The as-exfoliated WS$_2$ nanosheets contain high concentration of strained 1T phase metallic regions. To investigate the influence of strain and 1T phase on the catalytic activity, the monolayered WS$_2$ thin film was incrementally annealed and the catalytic activity after each thermal treatment was measured. This allowed evaluation of catalytic properties of the as-exfoliated WS$_2$ nanosheets without changing their dimensions or geometrical features. The HER characteristics are strongly linked to both the 1T phase concentration and strain as indicated by a gradual decrease in the exchange current density with increasing 2H fraction and decreasing strain, as shown in Figure 2d. We have also monitored the HER properties as a function of the proton concentration by varying the pH. The results provided in the Supplementary Information show that above a pH of 0.6 the overpotential and Tafel slope increase, consistent with a decrease in the concentration of H$^+$.

An important parameter for viability of a HER catalyst is the electrochemical stability. To investigate stability under electrocatalytic operation, we have measured the HER characteristics of $WS_2$ catalyst nanosheets by running over 10,000 cycles and monitoring the current density during continuous operation at $-0.3V$ for more than 100 hours. The current density *versus* time data provided in the Figure 3a shows that the values remain stable after a slight initial decrease (observed for all electrodes including Pt). The variation of the overpotential and concentration of the 1T phase as a function of time are shown in the insets of Figure 3a. It can be seen that the overpotential does not vary significantly with time after 120 hours as indicated by the fact that the 1T phase concentration remains stable during the electrochemical long term stressing, as indicated by Raman spectroscopy and XPS (Figure S12). Overall, these electrochemical results demonstrate that as-exfoliated $WS_2$ sheets could be effective electrocatalysts for HER. We have also investigated the impedance of the electrodes and found that it does not change dramatically with addition of 1T or 2H nanosheets to glassy carbon (Figure S14). The measurements show that the resistance of the system is slightly lower for the 1T nanosheets on glassy carbon electrode, suggesting that in addition to better catalytic activity, charge transfer is also facilitated.

To further investigate the stability of the 1T phase, we have carried out first principles calculations to probe the activation energy for transformation from the 1T to the 2H phase. Specifically, we carried out the nudged elastic band calculations to determine the energy barrier of phase transformation between 1T- and 2H-$WS_2$. We show that although the 2H-$WS_2$ is more stable with total energy 0.537eV lower than that of the 1T-$WS_2$, there is a energy barrier of ~ 0.95eV for phase relaxation, as indicated in Supplementary Information. The calculated energy barrier is consistent with experimental measurements of activation energy (~ 0.87eV) for 1T to 2H transformation [34]. The results are also consistent with our observations that the 1T phase remains constant in laboratory-stored samples for over 200 days (Figure S13).

Finally, we performed first principle calculations to obtain fundamental insight into mechanisms responsible for the very low overpotential and high TOF values for hydrogen evolution in 1T WS$_2$. The first step in HER is the adsorption of hydrogen onto the catalyst which can be written as $H^+ + e^- + ^* \rightarrow H^*$, where $^*$ denotes a binding site. The subsequent step is the release of molecular hydrogen by either $2H^* \rightarrow H_2 + 2^*$ or $H^+ + e^- + H^* \rightarrow H_2 + ^*$ reactions. This process has been theoretically modeled by density functional calculations and corroborated with experimental results [1,3,4,6]. It is known that a material is a good catalyst when the free energy of adsorbed atomic hydrogen is close to thermoneutral, *i.e.* $\Delta G_H \approx 0$. This is attributed to the fact that if the hydrogen does not efficiently bind to the catalyst or if it forms a strong bond then the proton/electron-transfer step and hydrogen release, respectively, will be inefficient, decreasing catalytic activity. We have calculated the free energy of atomic hydrogen adsorption of distorted 1T-WS$_2$ monolayers as a function of strain. Our calculations show that strain can significantly influence the free energy of atomic hydrogen adsorption on the surface of distorted 1T-WS$_2$ as shown in Figure 3b. Without strain, the free energy of H adsorption is 0.28 eV. However, the application of tensile strain leads to the enhancement of density of states near the Fermi level that facilitate hydrogen binding (Supplementary

Information). As shown in Figure 3b, the free energy is close to thermoneutral when the strain is varied between 2.0% - 3.0% strain. The free energy equal to zero can be extrapolated for a strain value of 2.7%. These results clearly indicate that the 1T $WS_2$ phase is catalytically active. For comparison, we have also calculated the effect of strain on catalytic activity of 2H $WS_2$ and found no improvement in catalytic activity. For example, even with strain as high as 4%, the free energy remained close to ~ 2eV, suggesting a very large thermodynamic barrier to HER activity. Furthermore, compressive stress did not have a significant influence on the density of states. The above experimental and theoretical results suggest that the presence of strain, as confirmed by our HAADF-STEM analysis, is an important factor in enhancing catalytic activity of $WS_2$ nanosheets. This combined with the scalability of the solution-based synthesis suggest that earth abundant as-exfoliated 1T $WS_2$ nanosheets could be potentially interesting catalysts for HER.

**Methods**

Solutions of tungsten disulfide nanosheets were prepared by immerging intercalated $WS_2$ powder in water. Bulk $WS_2$ powder was intercalated with lithium by reacting bulk $WS_2$ powder (0.3 g, Alfa Aesar) with n-butyllithium in hexane (1.6 M, 4 ml, Sigma-Aldrich). The process is comparable to that reported for $MoS_2$ in Ref. 25. More details are provides in SI.

HAADF STEM imaging was performed using JOEL JEM-2100F TEM/STEM with double spherical aberration (Cs) correctors (CEOS GmbH, Heidelberg, Germany). The acceleration voltage was 120 kV and the collecting angle was between 100 and 267 mrad. Peak pair analysis to obtain the stress tensor map was conducted with DigitalMicrograph software.

Electrochemical measurements were carried out with a 3-electrode cell using a VersaStat 3 potentiostat from Princeton Applied Research. Linear sweep voltamperometry with a 5 $mV.s^{-1}$ scan rate was performed in 0.5 M $H_2SO_4$ electrolyte de-aerated with hydrogen, nitrogen, or Ar with a saturated calomel electrode (CH Instruments) as the reference electrode, a platinum wire (Alfa Aesar) as the counter electrode and a glassy carbon electrode (CH Instruments) as the working electrode. The reference electrode was calibrated for the reversible hydrogen potential using platinum wire as working and counter electrodes in the electrolyte solution saturated with $H_2$. In 0.5 M $H_2SO_4$, E (RHE) = E (SCE) + 0.213 V. The site density measurements were carried out using the under potential deposition (UPD) method of Green et al. [Ref 40]. AC impedance measurements were performed in the same configuration at + 0.1 V from $10^6$ to 0.85 Hz with an AC voltage of 5 mV. Samples were directly deposited or transferred after annealing on the working electrode. 10 µl of the solution of exfoliated 1T-WS$_2$ was loaded over the glassy carbon electrode (equivalent to 0.1 – 0.2 µg-cm$^{-2}$ or approximately one continuous layer of WS$_2$ over the 1cm$^2$ electrode surface area) and dried. Then 7 µl of 5% Nafion solution in ethanol was drop casted on top to protect the WS$_2$ film. To prepare 2H-WS$_2$ samples, exfoliated WS$_2$ were annealed first prior to the transfer on the working electrode. See SI for more details.

The density functional calculations were carried out by using the Vienna *ab initio* simulation package (VASP) [43], with exchange-correlation functional described by Perdew-Burke-Ernzerhof generalized gradient approximation (PBE-GGA) [44] and interaction between core electrons and valence electrons by the frozen-core projector-augmented wave (PAW) method [45]. An energy cutoff of 600eV was used for plane wave basis expansion. More details are given in SI.

**Supplementary Information**

**Acknowledgements**

MC acknowledge financial support from NSF DGE 0903661. DV acknowledges from the Rutgers. HY acknowledges financial support for JSPS. GE acknowledges financial support from NRF Singapore. JL and VBS acknowledge support from Army Research Office through Contract W911NF-11-1-0171. TA acknowledges financial assistance from NSF (CAREER CHE-1004218, DMR-0968937, NanoEHS-1134289, NSF-ACIF, and Special Creativity Grant). RS and DCBA acknowledge the Coordenação de Aperfeiçoamento de Pessoal de Nível Superior, Brazil for fellowships. RS also acknowledges the Fulbright Agency, USA for financial support. TJ and MWC acknowledge partial support from JST-PRESTO.

**Author Contributions**

MC conceived the idea, designed the experiments, analyzed the data, and wrote the MS. DV conceived the idea and designed the experiments with MC, synthesized the $WS_2$ nanosheets, characterized them with AFM, Raman, and XPS, performed the HER measurements and analyzed the data. HY assisted in the synthesis and characterization of materials. JL and VS performed the theoretical work. TF and MWC performed the TEM work. RS, DCBA, and TA assisted DV with the HER measurements. GE analyzed the TEM and strain data as well as editing the MS. All the authors have read the MS and agree with its content.

**Additional Information**

The authors declare no competing financial interests.

**Figure Captions:**

**Figure 1 | Structure of chemically exfoliated WS$_2$.** a, AFM image of individual exfoliated nanosheets of WS$_2$. b-c, High resolution STEM images of as-exfoliated WS$_2$ monolayer showing regions exhibiting (b) the 1T superlattice and (c) 2H structures. The inset in (b) shows the strain tensor map generated from the STEM-HAADF image using peak pair analysis. Light (yellow) and dark (black) colors correspond to regions where the strain is in tension and compression, respectively.

**Figure 2 | HER electro catalytic properties of exfoliated WS$_2$ nanosheets.** Thin films of thicknesses of greater than five monolayers of as-exfoliated 1T and 2H phases of WS$_2$ nanosheets were deposited onto glassy carbon electrode. a, Polarization curves of bulk and as-exfoliated WS$_2$ (as-deposited film of 1T phase, sub-monolayer as-exfoliated film, and 2H phase after annealing at 300$^o$C) along with those corresponding to Pt nanoparticles and bulk WS$_2$ powder for comparison. b, Tafel plots obtained from the polarization curves show significant improvement in the Tafel slope due to exfoliation (60 mV/decade), whereas 2H and bulk samples exhibit much higher values of 110 mV/decade or higher. c, Turn over frequency (TOF) as function the overpotential for the different materials. The TOF reaches 175s$^{-1}$ at a potential of – 288 mV for as synthesized WS$_2$. d, Catalytic activity as a function of the 1T phase concentration obtained by annealing in an inert atmosphere.

**Figure 3 | 1T phase stability and free energy diagram for hydrogen evolution at equilibrium (U = 0) with tensile strain in atomically thin 1T WS$_2$.** a, The variation in current density versus time (up to 120 hours corresponding to >10,000 cycles) of 1T WS$_2$

electrode operation showing that the current density remains constant over the tested period. The insets in (d) show the percentage of change in overpotential (left inset) and variation in 1T phase concentration with time of electrode operation. b, The free energy of $H^+ + e^-$ is by definition equal to that of $1/2H_2$ at standard conditions (1 bar of $H_2$ and pH = 0 at 300 K). The free energies for the intermediate adsorbed state $H^*$ are calculated by using density functional theory and corrected for zero point energies and entropy. The hydrogen coverage of 1/16 S atoms is used for the calculations. The strain 2.7% for zero free energy is obtained by assuming a linear relation between free energy and strain.

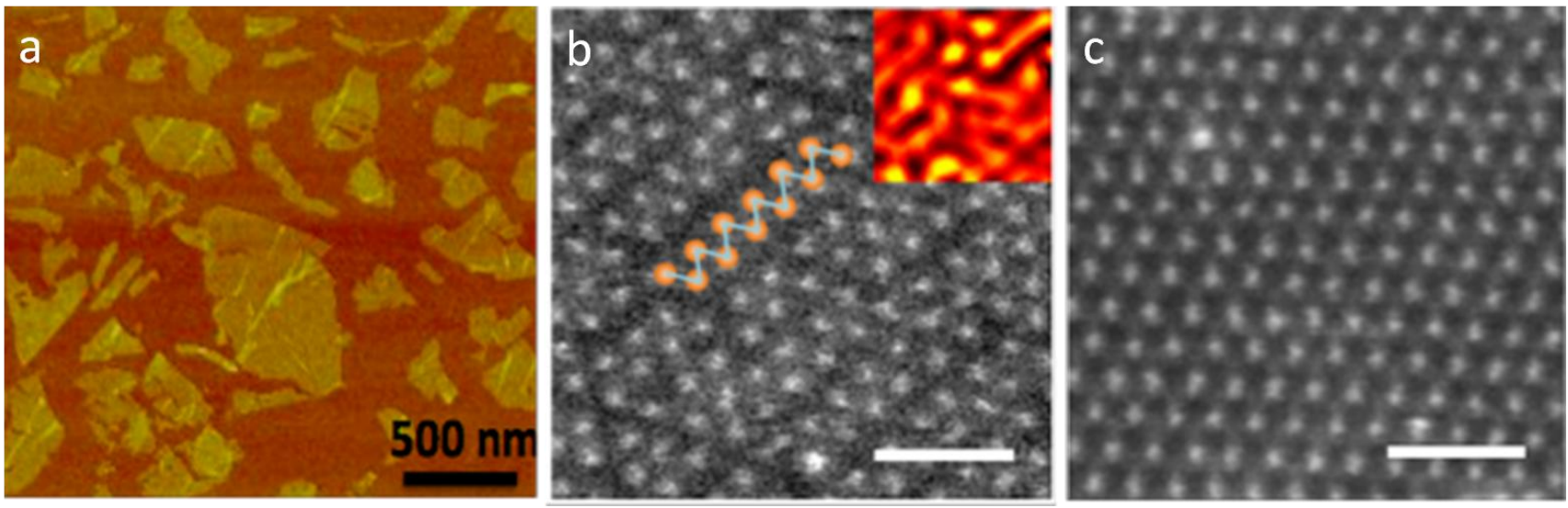

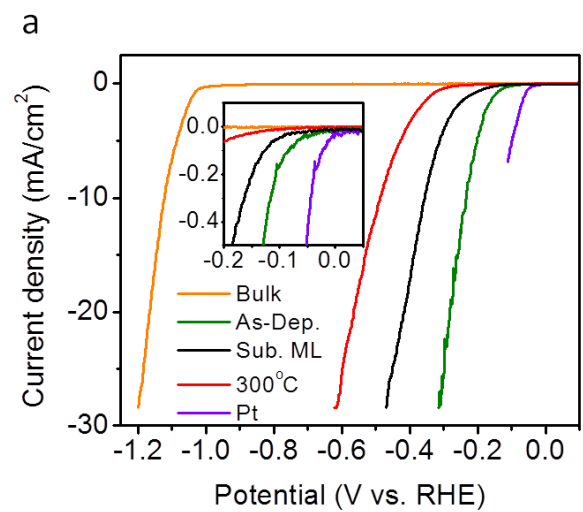
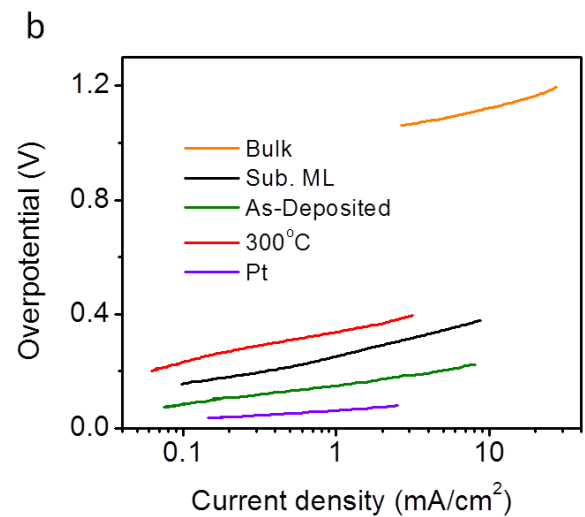
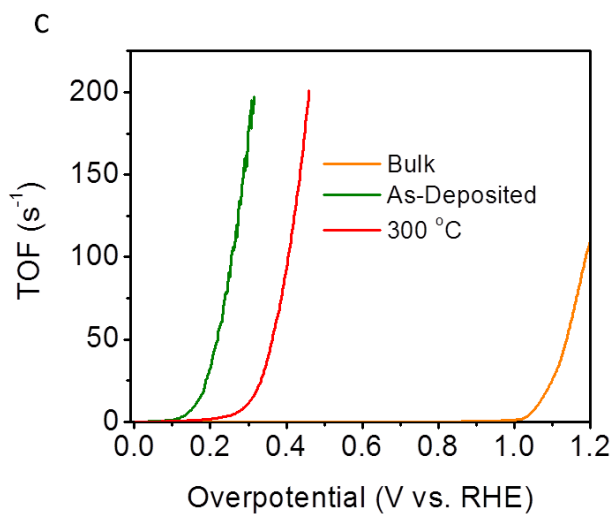
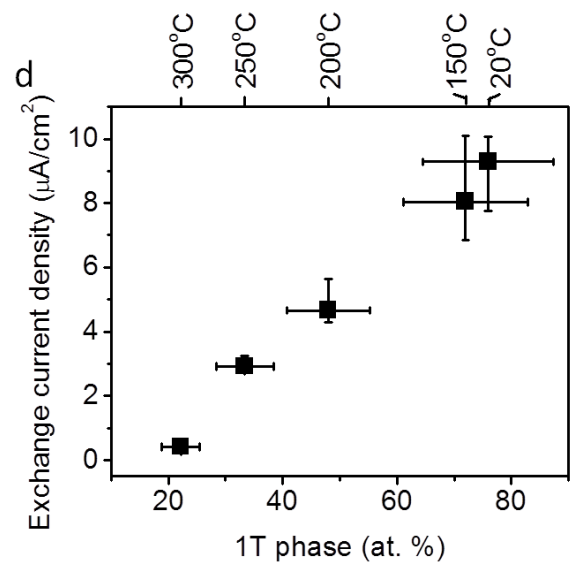

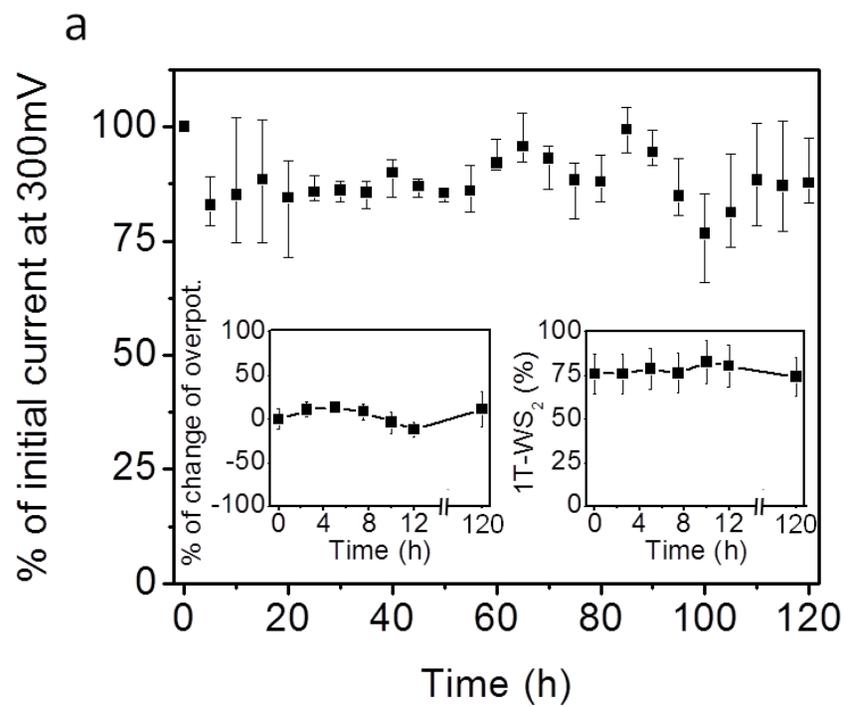 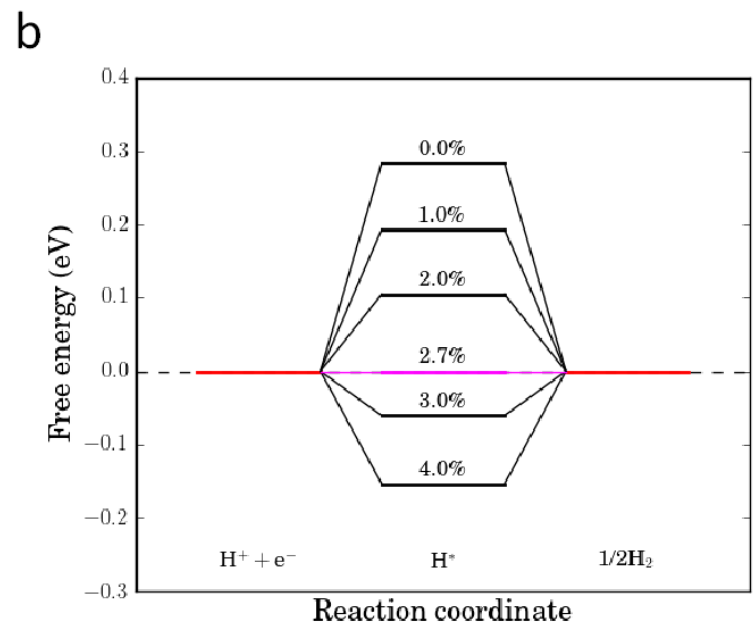